# Limitation in velocity of converging shock wave

Sergey G. Chefranov [*), Yakov E. Krasik and Alexander Rososhek

Physics Department, Technion, Haifa 32000, Israel

[*) csergei@technion.ac.il

The commonly applied self-similar solution of the problem of the converging shock wave (shock) evolution with constant compression of the medium behind the shock front results in an unlimited increase of the medium velocity in the vicinity of the implosion. In this paper, the convergence of cylindrical shocks in water is analyzed using the mass conservation law, when the water compression behind the shock front is a variable. The model predicts a finite range of radii, which depends on the adiabatic index of water and where the increase in pressure exceeds the sum of the change of the kinetic and internal energy densities behind the shock front. In this range of radii only the finite increase of the shock and water flow velocities is realized.

**I. Introduction**

Research studies of the evolution of converging spherical and cylindrical shocks are important both for fundamental and applied research problems related to the behavior of matter under extreme conditions, controlled thermonuclear fusion, and in astrophysics [1], [2]. In this research, the self-similar shock implosion approach is frequently used [3]. This approach considers an adiabatic ideal gas with a polytropic index $n > 1$ and maximal constant compression $\delta = (\rho_1/\rho_0) = (n+1)/(n-1)$ behind the shock front. Here $\rho_1$ and $\rho_0$ are the densities of the gas behind the shock front and at normal conditions, respectively. This approach does not consider the radiation losses of the compressed medium and the shock instabilities or the dynamics of the piston that generates this shock. This model yields an unlimited shock velocity and, consequently, pressure in the vicinity of the implosion. For example, in [4], using the self-similar approach for converging spherical and cylindrical shock implosions, the dependence $D_2 = D_1(R_2/R_1)^{-(1-\alpha)/\alpha}$ was determined. Here, $R(t) = A(-t)^\alpha; D = dR/dt = -\alpha A^{\frac{1}{\alpha}} R^{-\frac{1-\alpha}{\alpha}}$ is the shock radius and its front velocity, $\alpha = 0.6, 0.75$ are the self-similarity parameters for spherical and cylindrical shocks, respectively, and indices "1" and "2" correspond to the radii of convergence $R_1$ and $R_2 < R_1$ (see Fig.1). Thus, for given values of $D_1$ at $R_1$, the dependence $D_2(R_2)$ is calculated.

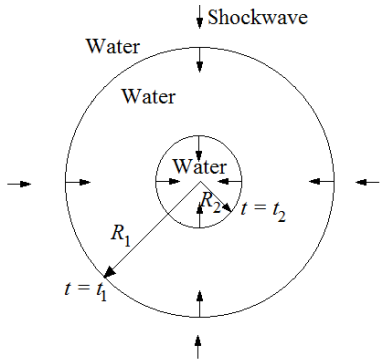

Fig. 1. Converging cylindrical shock in water with shock front at two time instants $t = t_1$ and $t = t_2$ corresponding to radii $R_1$ and $R_2$, respectively.

The self-similar approach describes the evolution of shock implosion velocity in the range of radii without considering information associated with the separation of the shock from the piston. However, when the shock has not yet separated from the piston, the shock evolution is determined by the piston velocity [3].

The evolution of the shock in the intermediate range of radii, i.e., the range where, despite a continuing increase in the distance between the shock and piston, the parameters of the shock can be influenced by its parameters shortly after its separation from the piston, remains unclear in experiments with electric underwater explosions[5]-[7].

In our study the variability of compression $\delta$ is taken into account when a description of the converging cylindrical shock evolution in water in this intermediate range of radii is presented. At the same time, it is the finite value of the rate of this variability of compression during the compression that can serve as the basis for establishing a quantitative assessment of that intermediate interval of compression radii. In the vicinity of the implosion axis, when the classical self-similar mode is established, the value of this velocity should decrease to zero.

It is noted in [8] and [9] that the compressibility of a liquid should limit the cumulation effect. The effects of compressibility were considered in [3], [10], and [11], but the obtained self-similar limiting regimes do not allow one to find the limitation in the cumulation due to this effect. The energy dissipation by the shock considered in [9] for sufficiently large Reynolds numbers, does not prevent unrestricted cumulation, when a cavity collapses in an incompressible liquid.

In our model the influence of the water compressibility made on the unlimited increase in the flow velocity near the implosion axis, also does not lead to its finite value. Nevertheless, it will be shown that in the vicinity of the implosion axis, starting from a certain threshold radius, the increase in the pressure becomes smaller than the sum of the kinetic and internal energy densities behind the shock front. By this means the corresponding limit on the shock and water flow velocities is obtained.

Considering next in Section II, using the mass conservation law, the Rankine-Hugoniot relations, and the equation of state (EOS) of water, the pressure, the water compression, and the flow velocity behind the shock front are obtained versus the shock radius. In Section III, the range of radii, where the increase of pressure exceeds the increase in the sum of the kinetic and internal energy densities behind the shock front, is determined.

**II. Mass conservation law and quasi-self-similar shock wave convergence**

Let us consider the convergence of cylindrical shock in water with a front of $R_1$ at time $t_1$ and $R_2$ at time $t_2$ (see Fig. 1).

The densities of water behind the shock front at $t_1$ and $t_2$ are $\rho_1$ and $\rho_2$, respectively. Let us introduce the dimensionless parameters: $\delta_1 = \rho_1/\rho_0$; $\delta_2 = \rho_2/\rho_0$; $x = R_1/R_2$ and $y = \delta_2/\delta_1 = \rho_2/\rho_1$, where $\rho_0$ is the density of undisturbed water. The mass conservation law at the shock front[3] dictates that $U_a = D_a(\delta_a - 1)/\delta_a$, where $D_a$ is the shock velocity and $U_a$ is the velocity of the water flow behind the shock front. The index $a = 1, 2$ is related to time $t_1$ or $t_2$. Let us consider the continuity equation for the cylindrical case with azimuthal symmetry:

$$\frac{\partial \rho}{\partial t} + \frac{1}{r}\frac{\partial}{\partial r} r\rho U = 0 \qquad (1)$$

In the laboratory coordinate system, we consider a water layer bounded by radii $R_1$ and $R_2$ (see Fig. 1 herein). The shock propagates with velocities $D_1 = -|D_1|$ and $D_2 = -|D_2|$ at times $t_1$ and $t_2$, respectively. Let us integrate (1) over the layer bounded by radii $R_2$ and $R_1$ (see also Eq. (A.2) in Appendix):

$$\frac{\partial}{\partial t}\int_{R_2}^{R_1} dr\, r\rho(r,t) + R_1\rho(R_1,t)U(R_1,t) - R_2\rho(R_2,t)U(R_2,t) = 0 \qquad (2)$$

In [8] and [9], the collapse of a vacuum cavity in an incompressible medium when $\rho = \rho_0 = const$ was analyzed using the self-similar power law $U(r,t)r = q(t) = U(R_c(t))R_c(t)$ for the medium velocity $U(r,t)$ versus radius $r$ and the cavity radius $R_c = R_c(t)$. Here $q(t)$ is a time-dependent value only. In the case of incompressible medium also from (2) one obtains relation $U(R_2;t) = U(R_1;t)(R_1/R_2)$. This dependence, showed an unlimited increase in the velocity of the medium versus $x = R_1/R_2 \gg 1$.

Let us consider an analogue of the self-similar power-law [8], [9] for a compressible medium. In this approach, based on the mass conservation law (2), it is possible to obtain the power-law dependence of the medium velocity versus $x = R_1/R_2$. Integration of (2) over time $t = t_1, t \to \infty$, considering that the first term in (2) has vanished, since at $t_1$ and $t \to \infty$ the water density is $\rho = \rho_0$, gives analogue of the quasi-self-similarity case considered in [8] and [9], but for a compressible medium:

$$R_1 I_1 = R_2 I_2;$$
$$I_1 = \int_{t_1}^{\infty} dt \rho(R_1;t) U(R_1;t); \qquad (3)$$
$$I_2 = \int_{t_2}^{\infty} dt \rho(R_2;t) U(R_2;t)$$

Here, it is accounted that the medium velocity $U(R_2;t)$ at $r = R_2$ is zero at $t \leq t_2$. Let us consider the mass fluxes balance without contribution of the mass fluxes reflected from the axis. In order to estimate the values of integrals $I_1$ and $I_2$, we assume that the water velocity and density time-dependence can be represented as

$$|U(R_a, t)| = U_a \exp(-\gamma_a(t - t_a)) H(t - t_a); \qquad (4)$$
$$H(t - t_a) = \begin{cases} 1, t \geq t_a \\ 0, t < t_a \end{cases}.$$

$$\rho(R_a;t) = \rho_0 + (\rho_a - \rho_0) \exp(-\gamma_{\rho a}(t - t_a)); a = 1,2. \qquad (5)$$

Here $H(t - t_a)$ is the Heaviside unit step function. In this case, one obtains:

$$I_a = U_a \rho_a f_a; \quad f_a = \frac{1}{\delta_a}\left[\frac{1}{\gamma_a} + \frac{(\delta_a - 1)}{\gamma_a + \gamma_{\rho a}}\right] \qquad (6)$$

Let us note that in the case $\gamma_a \gg \gamma_{\rho a}$, $f_a = \gamma_a^{-1}$, when the integrals $I_a = U_a \rho_a \gamma_a^{-1}$. In Fig. 2 one can see the temporal evolution of the density and the flow velocity of water at different radii of the cylindrical converging shock generated by underwater electrical explosion of a wire array having initial diameter of 10 mm and 4 kJ energy deposited into the array within ~1.2 μs. These dependencies were obtained using one dimensional (1D) hydrodynamic simulation (HD) [12] coupled with SESAME EOS for water and wire material [13] (see also Appendix hereto).

This simulation is written in the Lagrangian coordinates and assumes both axial and azimuthal symmetries, which are reasonable for cylindrical or spherical geometries. The simulation grid is divided into water and wire cells with the latter being subjected to energy deposition and expansion which sets the system in motion according to the Lagrangian approach. Initial conditions for this simulation at time zero (the first step) are normal conditions for water and wire material with no deposited energy. The solved equations are those of mass/momentum/energy conservation. The boundary conditions are reflecting, that means that when shockwave reaches either side of simulation space, it reflects back. To run this 1D HD simulation, the experimentally measured power deposition and the measured shock time-of-flight or shock trajectory are used as the input parameter and the verification parameter, respectively. At the beginning of each time step, this code computes the material density and interpolates the internal energy density from the increment in the measured deposited energy density between the previous and the current during the preceding time step. It is assumed that the electrical energy delivered into the wires during an explosion transforms into the internal energy with some efficiency. This efficiency acts as a fitting parameter to adjust the simulated shock time-of-flight or shock trajectory to the measured one. After the density and internal energy are known, the pressure and the temperature are computed from the SESAME tables.

Using exponential fit (4) and (5) (see Fig. 2) for these dependencies, one obtains values of $\gamma_1 \leq 1.52\ \mu s^{-1}$ and $\gamma_{\rho 1} \leq 0.14\ \mu s^{-1}$ for the flow velocity and density, respectively, at radius $R_1$ = 4 mm. Thus, the approximation $\gamma_1 \gg \gamma_{\rho 1}$ can be considered as reasonable. The same concern applies to smaller converging radii, when also $\gamma_2 \gg |\gamma_{\rho 2}|$. In the general case, the ratio $\gamma_2/\gamma_1$ depends on the radii ratio $x$. Thus, one obtains from (6) in case $\gamma_a \gg \gamma_{\rho a}$ that the flow velocities ratio $U_2/U_1$ behind the shock front is:

$$\frac{U_2}{U_1} = \frac{x}{y(x)} F;\ F = \frac{\gamma_2(x)}{\gamma_1} \qquad (7)$$

Using exponential fit for the data shown in Fig. 2, one obtains for flow velocity $\gamma_2(x=10) \approx 5.77\mu s^{-1}$. Thus, it is possible to introduce an empirical relation $F(x) \cong x^\beta$; $\beta = \frac{\ln(\gamma_2(x)/\gamma_1)}{\ln x} \approx 0.58$ for $x = 10$. The selection $x = 10$ is based on the estimate (see Section III), where we showed that the energetically optimal compression, when one obtains excess of the potential energy above the sum of the kinetic and internal energies behind the shock front, is limited by the value of the radius $R_2 > R_1/10$, that is $x < 10$.

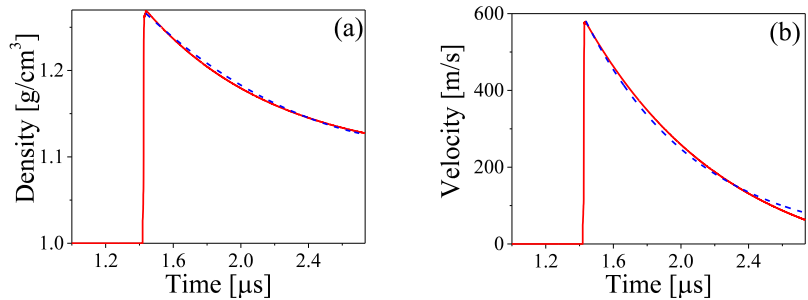

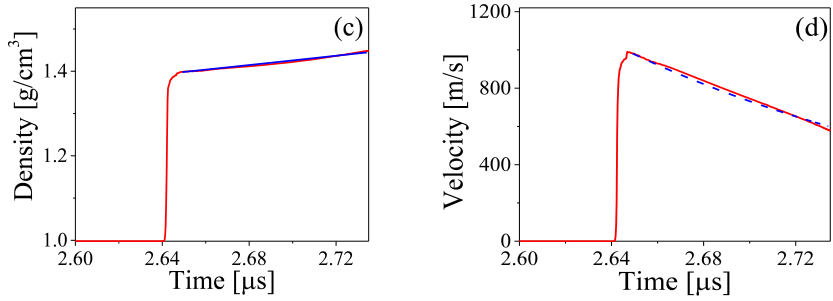
Fig. 2. Density and water flow velocity temporal evolution at the radius of 4 mm (a, b) and radius of 0.4 mm (c, d). Blue (dashed) lines are exponential fit using functions (4) and (5), where $t_1 = 1.437$ µs and $t_2 = 2.649 \mu s$. Red (solid) lines are the results of 1D numerical hydrodynamic simulations coupled with Sesame EOS of water.

To determine the form of function $y(x)$ in (7), let us consider this relation together with the Rankine-Hugoniot conditions $\rho_0 D^2 = \delta(p - p_0)/(\delta - 1); D = \delta(D - U)$ and the adiabatic EOS of water. From (7) and the Rankine-Hugoniot condition, the shock velocities ratio reads:

$$\frac{D_2}{D_1} = x \frac{(\delta_1 - 1)F}{(y\delta_1 - 1)} \qquad (8)$$

Applying the Rankine-Hugoniot condition, the ratio of pressures behind the shock front at radii $R_1$ and $R_2$ can be derived from (8) as:

$$\frac{p_2}{p_1} = x \frac{(\delta_1 - 1)F^2}{y(y\delta_1 - 1)} \qquad (9)$$

On the other hand, using the isentropic Tait's EOS for water (in the F. D. Murnaghan form) [14], $p_a - p_0 = (K_0/n)(\delta_a^n - 1)$, where $K_0 = 22$ kBar; $p_0 \approx 1$ Bar and $n \approx 7.15$, the ratio of pressures for $p > K_0/n >> p_0$ and $\delta_a^n >> 1$ is given as:

$$p_2/p_1 = y^n(x) \qquad (10)$$

Now, using (9) and (10) one obtains $y(x)$:

$$y^{n+1}(y\delta_1 - 1) = x(\delta_1 - 1)F^2 \qquad (11)$$

The dependence of the shock parameters on its radius is determined by relations (7) – (11).

Let us consider approximation $1 \gg 1/\delta_2(n+2)$, which is always satisfied, because $y > 1; n = 7.15$. Thus, the solution of (11) can be presented as:

$$y(x) = \left(\frac{xF^2(x)}{d}\right)^{\frac{1}{n+2}} \frac{1}{\left(1 - \frac{1}{y\delta_1}\right)^{\frac{1}{n+2}}} \approx \left(\frac{xF^2(x)}{d}\right)^{\frac{1}{n+2}} \left(1 + O\left(\frac{1}{(n+2)\delta_2}\right)\right)$$

$$d = \frac{\delta_1}{(\delta_1 - 1)}; F(x) = \frac{\gamma_2(x)}{\gamma_1} = x^\beta; \beta \approx 0.585$$

$$\frac{p_2}{p_1} \approx y^n \approx \left(\frac{xF^2(x)}{d}\right)^{\frac{n}{n+2}}; \quad \frac{U_2}{U_1} \approx (xF(x))^{\frac{n}{n+2}}\left(\frac{\delta_1}{\delta_1-1}\right)^{\frac{1}{n+2}} \qquad (12)$$

Here let us note that since $y(x) > 1$, the solution for $y(x)$ is valid only for $x > x_{min} = d^{\frac{1}{2(1+\beta)}}$. For example, for $\beta = 0.58; \delta_1 = 1.25$, one obtains the solution (12) applicability for $x > x_{min} = 1.66$. Let us define the shock convergence described by relations (12) as a quasi-self-similar approach. To avoid confusion with the common self-similarity parameter [15] for cylindrical shock $\alpha \approx 0.75$, we denote this parameter by $\alpha_q$. The value of the power in (12) can be represented

as $\frac{1}{\alpha_q} - 1$ resulting in

$$\alpha_q = \left(1 + \frac{(1+\beta)n}{n+2}\right)^{-1}$$

Thus, for the converging cylindrical shock in water one obtains $\alpha_q \approx 0.4$ for $\beta = 0.58$, which is smaller than in case of the self-similarity approach.

In order to compare (12) with the dependence for the water flow velocity obtained in the model for incompressible water [9], we will use similar, as in our model, relaxation of velocities at $R_1$ and $R_2$, namely $U(R_1; t) = U_1 \exp(-\gamma_1 t); U(R_2; t) = U_2 \exp(-\gamma_2 t); \gamma_2 = \gamma_2(x)$. After integration over time from zero to infinity, one obtains:

$$U_2 = U_1 \frac{\gamma_2}{\gamma_1} \frac{R_1}{R_2} = U_1 \left(\frac{R_1}{R_2}\right)^{1+\beta}; \frac{\gamma_2}{\gamma_1} = \left(\frac{R_1}{R_2}\right)^{\beta}; \beta = \ln\left(\frac{\gamma_2}{\gamma_1}\right) \ln^{-1}\left(\frac{R_1}{R_2}\right) \qquad (13)$$

Let us assume that in an incompressible medium, the indices of the velocity relaxation are close to $\beta = 0.58$ (see Fig. 2). In this case, one obtains the power-law dependence $U_2 = U_1 \left(\frac{R_1}{R_2}\right)^{1.58}$, which is shown in Fig. 3 together with the radial dependence of the velocity in case of compressible water. In the latter case, using (12) and $n = 7.15; \beta = 0.58$, the mass conservation law leads to the radial dependence for the flow velocity:

$$U_2 = U_1 \left(\frac{\rho_1}{\rho_1 - \rho_0}\right)^{\frac{1}{9.15}} \left(\frac{R_1}{R_2}\right)^{1.24}$$

One can see that the compressibility leads to a slower increase in the velocity of water which is related to the additional energy losses for water compression. However, the water compressibility does not eliminate an unlimited increase in the flow velocity.

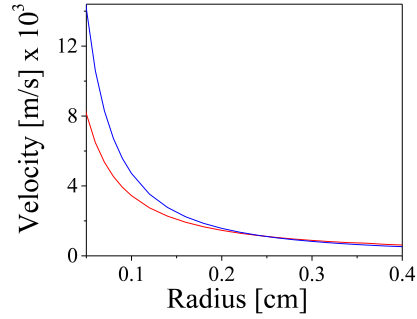

Fig.3. Dependencies of the of flow velocity vs. radius for compressible (red from (12) for $n = 7.15$ and $\beta = 0.58$) and incompressible (blue from (13) and [9]) cases.

The value of power, 1.24, in (12) is significantly (~3.76) larger than the value of power, (0.33) in common self-similar solution [3], which considers constant entropy of ideal gas compressed to its maximal value behind the shock front. Here let us note that the geometric factor, which is necessary to account to satisfy the integral law of conservation of mass (2) and (3), leads to the value of exponent power larger than 1, as in a cylindrical case of an incompressible water considered in [9]. Thus, some additional studies related to how the self-similar solution[3] satisfies the integral form of the mass conservation law, are required.

The quasi-self-similar solution (12) does not consider energy dissipation processes behind the shock front, which are realized in experiments and which are accounted in hydrodynamic simulations [12] coupled with EOS for water [13]. To account the energy dissipation in our model, we will use the results of numerical simulations of the water flow velocity relaxation (see Fig. 2 herein). In this case, the flow velocities ratio instead of (12) can be presented as:

$$\frac{\widetilde{U}_2}{\widetilde{U}_1} = \frac{U_2}{U_1} exp\bigl(-\gamma_1 x^\mu (t_2(x) - t_1)\bigr) \qquad (14)$$

Here, differing to relation (4), where the flow velocity temporal relaxation is related to a fixed radius, the ratio (14) determines the process of the energy dissipation during the shock propagation from larger to smaller radii. Here we assume that an exponential factor in (14), responsible for energy dissipation, does not prevent the use of the isothermal EOS for water and the Rankine-Hugoniot relations for ratio $U_2/U_1$ determined by (12). In Fig. 4(a), the dependence $\widetilde{U}_2/\widetilde{U}_1$ [see (14)] versus ratio of radii $x$ is shown. Here values of $\mu(x)$ [see Fig. 4(b)] are determined using the fit of $\widetilde{U}_2/\widetilde{U}_1$ with the results of one dimensional hydrodynamic numerical simulations of the shock convergence for the same conditions as in Fig. 2 given herein. In Fig. 4(a) we also show the dependence of $\frac{U_2}{U_1} = f(x)$ which follows from (12). One can see that the quasi-self-similar solution (12), as expected, differs strongly from the numerical simulations and shows the necessity for energy dissipation accounting.

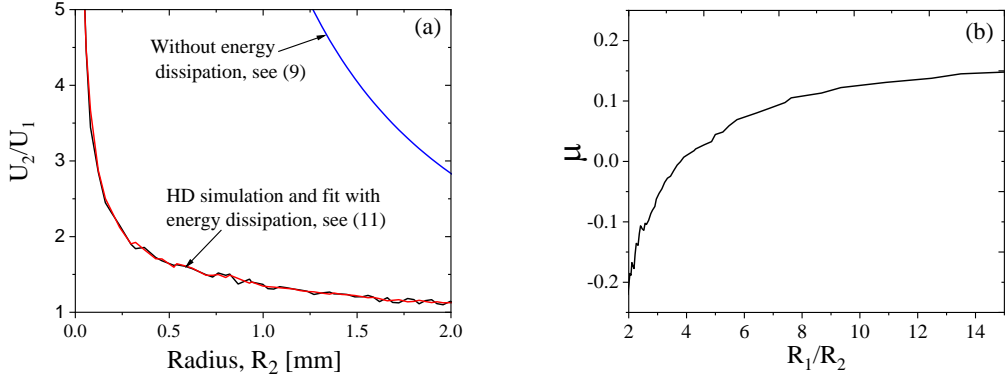

Fig. 4. Water flow velocities ratio vs. the radius of convergence (a) and values of μ vs. radii ratio (b).

In case $x \gg 1$, the value $t_2(x) - t_1 \to t_F - t_1 = const$, where $t_F$ is the time, when the shock approaches the implosion axis. Thus, in the vicinity of implosion, an increase in energy dissipation occurs solely due to the increasing of $x^\mu$ in (14), thus limiting the increase in water flow velocity. Indeed, in the limit $x \gg 1; t_2 \to t_F \approx 2.73$ µs (see Fig. 2) and using (12), the ratio (14) reads:

$$\frac{\tilde{U}_2}{\tilde{U}_1} = A x^a \exp(-B x^\mu);$$

$$A \approx 1.19; a = 1.23; B = 2.02; \mu = 0.125 \qquad (15)$$

This flow velocities ratio has its maximum at $x_{d\_max} = \frac{1}{\mu} ln\left(\frac{a}{B\mu}\right) \approx 12.7,$ which determines the maximal radius to which water velocity increases behind the shock front. In [9], a similar effect of the limited energy accumulation was obtained due to the viscosity effects for the Reynolds numbers smaller than a threshold value.

### III. Energy-optimal compression range of converging radii

In this Section, it will be shown that even without considering the energy dissipation, there is a restriction on the range of radii $x < x_{max}$ related to a violation of the energetically optimal compression mode. Using relations (7) – (12), one can determine this range of radii, when the increase in the pressure exceeds the increase in the sum of the kinetic and internal energy densities of the water flow. The average kinetic energy density transferred by the shock to the water flow behind its front during $\Delta t = t_2 - t_1$ can be defined as $\Delta E_K = \rho_2 \frac{U_2^2}{2} - \rho_1 \frac{U_1^2}{2}$. Using the Rankine-Hugoniot relations, the change in kinetic energy density reads

$$\Delta E_K = \frac{\rho_1 U_1^2}{2}\left(y \frac{U_2^2}{U_1^2} - 1\right) \equiv \frac{p_1}{2}\left(\frac{p_2}{p_1}(y\delta_1 - 1) - \delta_1 + 1\right) \qquad (16)$$

The change in the internal energy density $\Delta E_I = E_2 - E_1$, calculated using (7) and (9), when $\rho_1 U_1^2 = p_1(\delta_1 - 1)$ and the Rankine-Hugoniot condition [3], [16] $E_a = \varepsilon_a \rho_a, \varepsilon_a - \varepsilon_0 = \frac{(p_a + p_0)}{2}(\frac{1}{\rho_0} - \frac{1}{\rho_a})$ for $p_a \gg p_0, E_a \gg \delta_a E_0$, is found to be equal to $\Delta E_K$ where $a = 1; 2$. Let us note that the value of $\Delta E_I$ is obtained using only the Rankine-Hugoniot condition. In the general case, the value of $\Delta E_I$, in addition to the thermal component associated to the thermal motion of particles, should also include a change in the potential energy of the interaction of these particles (see Eq. 8.4 in [3]). Here let us note that the equality $\Delta E_I = \Delta E_K$ agrees with the data presented in Ref. [17] regarding shocks in water where also is shown that the condition $\Delta E_P > \Delta E_I + \Delta E_K$ is valid only for pressures below some threshold value.

The change in the pressure during $\Delta t$ can be written as follows:

$$\Delta E_P = p_1 \left(\frac{p_2}{p_1} - 1\right) \qquad (17)$$

Now, let us estimate the minimal shock radius for which the energy compression ($\Delta E_P > \Delta E_I + \Delta E_K$) is optimal. Below this radius, the increase in the sum of the change in the kinetic and internal energy densities exceeds the increase in the pressure change; i.e., $\Delta E_K + \Delta E_I = 2\Delta E_K > \Delta E_P$. In the latter case, for instance, hydrodynamic instability may occur, leading to the appearance of swirling vortex motion (see [9], [18] and [19]), which can change the water flow's radial convergence. Using (16) and (17), condition $\Delta E_P > 2\Delta E_K$ reads

$$y^n(2 - \delta_1 y) > 2 - \delta_1 \qquad (18)$$

One can see that, for $\delta_1 < 2$, the condition $y < y_{max} = 2/\delta_1$ should be satisfied. Thus, for instance for $\delta_1 = 1.25$ inequality (18) is valid for $1 < y < y_t = 1.6$, when $y_t \approx y_{max} = 2/\delta_1$. Using (12) for $y$, the range of radii within which $\Delta E_P > 2\Delta E_K$ is valid can be obtained as

$$x < x_{max} = \left(\frac{2^{n+2}}{(\delta_1 - 1)\delta_1^{n+1}}\right)^{\frac{1}{2(1+\beta)}}$$

$$R_2 > R_{2min} = R_1/x_{max} \qquad (19)$$

Applying (19), for instance for $\delta_1 = 1.25$, one obtains $R_{2min} \approx 0.15 R_1$, which is almost twice the radius $x_{d\_max}$ obtained in Section II herein. Thus, when condition (19) is satisfied, an energetically optimal compression can be realized.

Finally, using (12) and (19), one can estimate that, for the quasi-self-similar mode, the maximum pressure when the shock approaches $R_{2min}$ is

$$p_{2max} \approx p_1 \left(\frac{2}{\delta_1}\right)^n = \frac{K_0}{n}(\delta_1^n - 1)\left(\frac{2}{\delta_1}\right)^n \approx 2^n \frac{K_0}{n} \qquad (20)$$

For implosion in water, for $n = 7.15$, the maximal pressure is $p_{2max} = 440 \, kBar$. One can see that the maximum pressure value in the vicinity of implosion is independent of the initial compression of the converging shock and its value depends only on the adiabatic index, which in general depends on the pressure [14].

### IV. The scope of applicability of the results obtained

Based on the mass conservation law, the model for converging cylindrical shock waves is suggested. The suggested model considers the conservation laws in the form of the Hugoniot relations and the integral form of the mass conservation law. The limits of applicability of these laws relate to dissipation processes, which are not accounted in our basic model. In addition, we use the isothermal EOS for water which applicability is limited by pressures of a few hundreds of kBar. However, a more significant limitation actually applies to the value of the index *n* in the EOS, which is considered constant. In general case, the value of *n* depends on temperature and, respectively, on pressure and density.

## V. Summary

The model that predicts the finite energy-optimal range of the radii of the converging shocks is represented herein. This mode is characterized by the dominance of the pressure increment in the compression process, compared with the increment of the kinetic energy of the motion of the medium behind the shock wave front, associated with the development of instability of purely radial motion of the medium. In this model of implosion an ultra-high, but however limited pressures can be obtained behind the shock front. The possibility of linking the condition of energetically optimal compression with the problem of stability of the converging shock wave front is noted.

## Acknowledgments

We thank A. Velikovich, J. Leopold, V. Gurovich, A. Virozub, D. Maler and S. Bland for fruitful discussions and comments.

This research was supported by the Israeli Science Foundation Grant No. 492/18.

## Appendix (see also [20])

Since it is difficult to measure the water parameters in the vicinity of the exploding wire array along with pressure and temperature, the experimental results are compared with one-dimensional (1D) hydrodynamic (HD) simulation. Using mass, momentum and energy conservation, the hydrodynamic equations, written in Lagrange mass $m$ (on the unit of length) coordinates, derived for the cylindrical case, have the following form [20]-[22]:

$$V(m,t) = \frac{\partial r(m,t)}{\partial t}; \qquad (A.1)$$

$$\frac{\partial r^2}{\partial m} = \frac{1}{\pi \rho(m,t)} \qquad (A.2)$$

$$\frac{\partial V}{\partial t} = -2\pi r \frac{\partial p(m,t)}{\partial m} \qquad (A.3)$$

$$\frac{\partial \varepsilon(m,t)}{\partial t} = -p \frac{\partial}{\partial t}\left(\frac{1}{\rho}\right) \qquad (A.4)$$

$$\varepsilon = \varepsilon(T,\rho); p = p(T;\rho) \qquad (A.5)$$

Here, $\rho$ and $T$ are the density and temperature of the material, $p$ and $\varepsilon$ are the pressure and the specific energy of the material, and $V$ is the radial component of the velocity. The mass coordinate $m$ is defined as $m = 2\pi \int_0^r dr_1 r_1 \rho(r_1;t)$. This model [21] approximates a converging cylindrical piston. Taking the simulated volume to be smaller than the array's half width, one can neglect changes in the shock profile arising from boundary effects. In practice, we take the radius of the cylindrical array to be large, and we look at small distances from the array where the converging shock can be treated as a planar shock. This model can also be considered (with some geometrical adjustments) for the spherical case. The simulation is solved using a cross scheme [22] coupled with the SESAME [13] EOS for water, Cu and Al along with the deposited energy into the wires which is obtained experimentally. The radiative heat transfer and heat conduction are neglected as these effects are negligible. This system of equations needs to satisfy several boundary conditions. First, in the vicinity of the implosion onset ($r = 0$) and at $r = R_{max}$ (the upper limit of the radius taken in the simulation,) the boundary conditions are given as:

$$V = 0; \frac{\partial p}{\partial r} = 0; \frac{\partial \rho}{\partial r} = 0; r = 0, r = R_{max} \tag{A.6}$$

Moreover, the velocity between two neighboring layers, made of different materials, is equal such that

$$V^{water}(r_n) = V^{Cu}(r_n) \tag{A.7}$$

In (A.7) $V$ is the velocity of the relevant material, and $r_n$ is the location of the relevant cell. The spatial index $n$ denotes the location of each cell on the grid. This condition prevents the mixing of different materials. Using the cross scheme, the discretization of the equations yields the following set:

$$\tilde{r}_n = r_n + V_n \tau; \tilde{V}_n = V_n - \tau \frac{2\pi r_n}{m_n}(p_n - p_{n-1}); \tilde{\rho}_n = \frac{m_n}{\pi((\tilde{r}_{n+1})^2 - (\tilde{r}_n)^2)}; \tilde{\varepsilon}_n = \varepsilon_n - p_n\left(\frac{1}{\tilde{\rho}_n} - \frac{1}{\rho_n}\right) \tag{A.8}$$

Here $\tau$ is the time step $n$ is the spatial index and $m_n = \pi \rho_n (r_{n+1}^2 - r_n^2)$ is the mass of the layer $n$. The 'tilde' sign denotes the next time step. The simulation begins with setting initial conditions for the different materials $p_0, T_0, \rho_0, \varepsilon_0$. The upper limit for the radius $R_{max}$ is set along with the total grid spacing, which is defined using an integer $N$ such that $1 \leq n \leq N$. The time loop begins with calculating the velocity of the wall between two neighboring mass cells. Now, the coordinates for the new cell are determined using the first equation in (A.8). Next the density and the internal energy are calculated. To determine the pressure and temperature as a function of density and internal energy, the SESAME EOS database is employed. Since the pressure, used in (A.8), is of the previous time step, the calculation of density and internal energy, followed by the extraction of the temperature and pressure, is performed twice to improve accuracy. Lastly, external energy is deposited utilizing the experimentally obtained data of the current and resistive voltage. The external energy is deposited to the metal layers according to

$$\tilde{\varepsilon}_n = \varepsilon_n + \tau \frac{\alpha P L_{wire} m_n}{M_{wires}^2} \quad (A.9)$$

Here the power $P = V_R I$ determined by the product of the resistive voltage $V_R$ and the measured current ($I$) and $M_{wires}$ is the total mass of the wires, which have lengths $L_{wire}$; $0<\alpha<1$ is a fitting parameter which allows us controlling the energy deposition such that the simulated results fit the experimental data. With this calculation the time loop begins again following the same algorithm for the next time step.

Note also that the theory presented in this paper is based on the use of the integral form of the law of conservation of mass in the form (2), which corresponds to a similar consideration of the continuity equation in the form (A.2) when using Lagrange variables.

Since in the simulation algorithm discussed above, all considerations are local in nature, therefore, at each integration step, the Euler hydrodynamic equations used for a compressible medium only locally take into account the curvature of the front and actually coincide with one-dimensional equations in unlimited space for which there is an exact Riemann solution [16]. In this regard, it is possible to develop an alternative stimulation scheme directly in Euler variables. This is feasible both by applying the Riemann solver method known in computational fluid dynamics [23] and the analytical explicit form of the Riemann solution obtained in [24].

## Data availability

The data that support the findings of this study are available from the corresponding author upon reasonable request.

...